\documentclass{article} 
\pdfoutput=1
\usepackage{arxiv}

\usepackage{times}
\usepackage{xspace}
\usepackage[dvipsnames,table]{xcolor}
\usepackage{hyperref}
\usepackage{latexsym}
\usepackage{tikz}
\usepackage{amsmath}
\usepackage{booktabs}
\usepackage[round]{natbib}
\usepackage{graphicx}
\renewcommand\cite{\citep}	%

\newcommand{\overreportingcrsdata}{ 75.35\xspace}

\newcommand{\caredataoverreported}{ 21.8 \xspace}
\newcommand{\caredataoverreportedclassifier}{ 54.14 \xspace}

\usepackage[T1]{fontenc}

\usepackage[utf8]{inputenc}

\usepackage{microtype}
\newcommand{\tabref}[1]{\tablename~\ref{#1}}

\hypersetup{
pdftitle={Using Text Classification with a Bayesian Correction for Estimating Overreporting in the Creditor Reporting System on Climate Adaptation Finance},
pdfauthor={Janos Borst, Thomas Wencker, Andreas Niekler},
pdfkeywords={Climate Finance, Development Aid, Climate Change Adaptation, Machine Learning},
}

\begin{document}

\title{Using Text Classification with a Bayesian Correction for Estimating Overreporting in the Creditor Reporting System on Climate Adaptation Finance }
\thispagestyle{empty}

 \author{Janos Borst\\Leipzig University\\janos.borst@uni-leipzig.de \And Thomas Wencker\\German Institute for Development Evaluation\\Thomas.Wencker@deval.org
\And Andreas Niekler\\Leipzig University\\andreas.niekler@uni-leipzig.de}

\maketitle

\begin{abstract}
Development funds are essential to finance climate change adaptation and are thus an important part of international climate policy. %
However, the absence of a common reporting practice makes it difficult to assess the amount and distribution of such funds. %
Research has questioned the credibility of reported figures, indicating that adaptation financing is in fact lower than published figures suggest. %
Projects claiming a greater relevance to climate change adaptation than they target are referred to as "overreported". %
To estimate realistic rates of overreporting in large data sets over times, we propose an approach based on state-of-the-art text classification. %
To date, assessments of credibility have relied on small, manually evaluated samples. %
We use such a sample data set to train a classifier with an accuracy of $89.81\% \pm 0.83\%$ (tenfold cross-validation) and extrapolate to larger data sets to identify overreporting. %
Additionally, we propose a method that incorporates evidence of smaller, higher-quality data to correct predicted rates using Bayes' theorem. This enables a comparison of different annotation schemes to estimate the degree of overreporting in climate change adaptation. %
Our results support findings that indicate extensive overreporting of $32.03\%$ with a credible interval of $[19.81\%;48.34\%]$. %

\end{abstract}
\xspace

\section{Introduction}
There is international consensus on the need to respond to the global threat posed by climate change (Paris Accord, Article 2). %
Development funds are essential to finance climate change adaptation and are thus an important part of international climate policy. %
The 2009 Copenhagen Accord \cite{unfccc2009copenhagen} aimed to mobilize USD 100 billion by 2020. %
Implementation of climate change adaptation measures is one of five targets set to reach the 13th Sustainable Development Goal (SDG): ``Take urgent action to combat climate change and its impacts''. %

The Creditor Reporting System (CRS), maintained by the OECD Development Assistance Committee (DAC), monitors adaptation finance flows from OECD DAC member countries to developing countries. %
One of the challenges in ensuring valid reporting - or at least comparable figures - across reporting agencies is that the agreements mentioned above lack indicators. %
To this end, the OECD DAC established in 2009 the Rio markers on climate change adaptation (CCA). %
For each aid activity, donors report whether it contributes to CCA, i.e. reducing ``the vulnerability of human or natural systems to the current and expected impacts of climate change, including climate variability, by  maintaining  or  increasing resilience, through increased ability to adapt to, or absorb, climate change stresses, shocks and variability and/or by helping reduce exposure to them'' \cite[p. 4]{oecd_dac_rio}. %
Activities are eligible for a marker if 
``a) the climate change adaptation objective is explicitly indicated in the activity documentation; and
b) the activity contains specific measures targeting the definition above.'' \cite[p. 4]{oecd_dac_rio}.
The Rio marker $r$ can take three values: 2, if CCA is the principal objective; 1, if CCA is a significant objective; and 0, if CCA is neither a principal nor a significant objective. \\
\indent However, there is increasing evidence that the level of adaptation financing is in fact lower than published figures suggest \cite{Weikmans2017AssessingTC, junghans_different_nodate}. %
One possible reason is that there is no common practice for reporting climate finance \cite{weikmans_international_2019, weikmans_what_2020} and reporting agencies thus follow different reporting rules. %
This makes it difficult to assess the total amount of CCA finance, to compare commitments between donors, and to assess the geographical and sectoral distribution of funding \cite{weikmans_international_2019}. %
Moreover, CCA finance estimates  vary among reporting agencies \cite{yeo_climate_2019}.
Hence, aggregate figures of adaptation finance are increasingly considered unreliable given that they comprise thousands of individual aid activity descriptions from the CRS data. %
Consequently, assessments of credibility have to date relied on analysis of small samples covering a limited period of time \cite{Weikmans2017AssessingTC}.

This study applies state-of-the-art machine learning methods to estimate overreporting of CCA finance for all aid activities as reported in the OECD DAC CRS since the introduction of Rio markers. %
Our main challenge in applying machine learning methodology is the quality and quantity of available annotated data. %
We have access to two data sets re-evaluated by experts and published in previous work: The first is small, but following a thorough re-evaluation process, we regard it as high quality. %
One concern is that the current de facto standard of fine-tuning language models tends to be unstable with very small data sets and is hard to evaluate properly. %
The second set is much larger, but because it was re-evaluated with access to less information, we regard it as lower quality; nevertheless, its size makes it adequate for training. %
We propose to combine these two data sets, using the larger for training and extrapolation, and the smaller, higher-quality set to correct first estimates. %
The contribution of this paper is two-fold:
\begin{enumerate}
    \item We propose and evaluate a machine learning model to detect overreporting in the CRS data and discuss extrapolation.
    \item We propose and attempt to use a Bayesian Framework for correction of extrapolated overreporting rates.
\end{enumerate}

\section{Related Work}%
In recent years, several studies have estimated the level of overreporting in CCA finance \cite{MICHAELOWA20112010, Weikmans2017AssessingTC, junghans_different_nodate, Schramek2021ADAPTATIONF}. %
These studies are distinguished, first, by the rigor of their methodology to assess overreporting and, second, by the number of aid activities they analyze. %
Some studies classify multiple aid activities but employ, rather simplistically, keyword searches only on short descriptions of aid activity \cite{MICHAELOWA20112010,roberts_reality_2008,junghans_different_nodate}. 
Other studies examine only a few aid activities by scrutinizing extant project documentation against in-country expert assessments \cite{Schramek2021ADAPTATIONF}.
Weikmans et al. (\citeyear{Weikmans2017AssessingTC}) strike a balance by manually assessing a large number of short project descriptions. \\
We are among the first to apply state-of-the-art machine learning - which allows us to code all aid activities reported in the OECD DAC CRS database - to fully automate the process of detecting overreporting of CCA finance.  %
Moreover, our method can be easily applied to future data releases of the OECD DAC CRS data as well as comparable text data. \\
Machine learning approaches to classify official development assistance are still rare.
\citet{OECD_SDG_2019} used machine learning to classify SDGs.
More recently, \citet{toetzke_consistent_2022} developed a machine-learning classifier to identify climate finance based on the title and descriptions of bilateral aid activities in the OECD DAC CRS dataset. ClimateFinanceBERT first classifies the relevance of aid activities to adaptation, mitigation, or the environment. Subsequently, relevant activities are further differentiated into ten categories. In contrast, our classifier directly predicts Rio Markers for climate change adaptation. Moreover, we address possible shortcomings due to the limited information contained in the OECD DAC CRS descriptions by integrating evidence from a high-quality re-evaluation. 
\\
Here, we rely on textual resources to automatically assign Rio markers to CRS Reports. %
Project reports typically contain both short and long descriptions of the project goals, ranging from one to a few sentences. %
Currently, neural networks produce virtually every state-of-the-art result in text classification, either by training task-specific architectures, e.g. \cite{kim_convolutional_2014} or 
adapting pre-trained language models to a given task \cite{devlin-etal-2019-bert, Liu2019RoBERTaAR, yang_xlnet_2019,aly_hierarchical_2019,Pal_2020}. 
Also, recent works have achieved both higher overall performance \cite{devlin-etal-2019-bert, yang_xlnet_2019} and greater sample efficiency, achieving better results with less data ~\cite{halder-etal-2020-task}. %
The typical problems of previous classical machine learning approaches in text classification, like out-of-vocabulary or ambiguity, are directly handled by the language model, which is especially important in this case, because not all of the texts we deal with are free of orthographic and syntactical anomalies.
We experiment with these models in various combinations to find the best fit for the task at hand.

\section{Automatic Classification of Climate Change Adaptation Markers}

\subsection{Data}
The CRS tracks OECD DAC member countries' aid activities. %
This study works with the original CRS data and two re-evaluated data sets:\\
\indent  \textbf{Creditor Reporting System (CRS)}: %
The publicly available CRS contains harmonized data on aid activities. %
We use CRS data from 2006 to 2019 containing 1,529,984 aid activities. %
It includes up to 91 fields of data for each aid activity. %
The most important information in our context is flagged by: donor, recipient, Rio marker, project title, and short and long description. \\
\indent \textbf{WK}: Weikmans et al. (\citeyear{Weikmans2017AssessingTC}) sampled 4,757 aid activities from 2012 CRS data and manually re-evaluated the Rio markers based on the aid activity descriptions. The re-evaluation includes a new marker (99) to indicate insufficient information for determining the Rio marker. %
\cite{Weikmans2017AssessingTC} argue that label 99 can be treated as 0 (not climate adaptation related) because the Rio marker methodology explicitly requires a CCA objective to be indicated in the aid activity documentation. \\
\indent \textbf{CARE}: Schramek and Harmeling (\citeyear{Schramek2021ADAPTATIONF}) of the Cooperative for Assistance and Relief Everywhere (CARE) sampled 117 aid activities from the CRS. %
Each case was re-evaluated and assigned a new Rio marker by experts with access to detailed project-level information beyond the data contained in the CRS. 

\subsection{Approach}
We consider CARE as a high-quality re-evaluation with very few samples. %
The WK data set has substantially more observations, but the re-evaluation had access only to CRS information. %
Since information from the CRS can be very limited, especially in cases where CCA is not the primary goal, this likely leads to a higher proportion of projects being considered overreported. However, the WK data set is substantially larger than the CARE data set and can be used to train a classifier, which is why we use WK for the training and CARE to estimate a correction factor to extrapolate the CARE annotations implicitly. \\
\indent Our approach is as follows: First, we train a high-quality classification model on the WK data set using information only from the CRS meta fields and the re-evaluated Rio markers. %
We mark a project as overreported if the classifier predicts a lower Rio marker than reported. %
Second, we calculate the classifier's overreporting rate on the CARE data set. %
By comparing overreporting with the high-quality re-evaluation, we can estimate an error factor between the two annotation schemes in a Bayesian framework. %
Finally, we extrapolate to the complete CRS database and estimate overreporting rates for both annotation schemes.

\subsection{Model Training and Model Selection}%

The WK data in the CRS provides us with text descriptions and the corresponding Rio markers (0, 1, 2 and 99), which we consider input and target of the classifier respectively. %
To find the best model, we test various language models with standard finetuning and in combination with a CNN architecture to find the best model. %
The CNN architecture follows \cite{kim_convolutional_2014} and comprises four 1D-convolutions with kernel sizes 3,4,5 and 6, with 100 filters each. %
The resulting vectors are max-pooled and projected by a linear layer onto the number of classes. \\
We conduct experiments with 
a RoBERTa \cite{Liu2019RoBERTaAR} base model, a BERT
\cite{devlin-etal-2019-bert} base model and a distilled version of RoBERTa \cite{Sanh2019DistilBERTAD} as
published in the Hugging Face 'transformers' library \cite{wolf-etal-2020-transformers}. %
We use the AdamW \cite{Loshchilov2019DecoupledWD} optimizer with a learning rate of $5e\text{-}6$, a batch size of 32, 25 epochs, and check-pointing to restore the best model with regards to average macro F1 score.
To ensure stability and quality, we test these hyper-parameters for all models with tenfold cross-validation to identify the best model. The results of the cross-validation experiments for all model combinations are shown in ~\tabref{tab:model_selection}. %
We used one Tesla V100 32GB for these experiments, one tenfold cross-validation for one model combination (one row in Table \ref{tab:model_selection}) took around five hours to complete for BERT and RoBERTa, and around 2.5 h for the distilled RoBERTa.
After ensuring an average performance, we randomly split the data 80/20 and train a model with the same parameters.\footnote{Code and final model will be made available upon publication to not jeopardize anonymity of the review.}%
\begin{table*}[h]   
\small
    \centering
\begin{tabular}{ll|llll}
\toprule
            &               &          accuracy & \multicolumn{3}{l}{macro} \\
            &               &                &                 P &                 R &                F1 \\
\hline
\hline
CNN & roBERTa-base & \textbf{89.81 $\pm$ 0.83} &   \textbf{84.83 $\pm$ 2.0} &  $80.54 \pm 2.03$ &   \textbf{82.31 $\pm$ 1.8} \\
            & BERT &   $88.89 \pm 1.3$ &  $83.35 \pm 2.13$ &  $79.64 \pm 2.81$ &  $80.97 \pm 2.28$ \\
            & distilroberta &   $89.16 \pm 1.3$ &  $83.47 \pm 2.24$ &   \textbf{81.28 $\pm$ 2.96} &  $82.12 \pm 2.21$ \\
Transformer & RoBERTa-base &  $89.64 \pm 1.61$ &   $84.1 \pm 2.33$ &  $80.58 \pm 3.74$ &   $82.0 \pm 2.87$ \\
            & BERT &   $89.27 \pm 1.5$ &  $84.57 \pm 3.39$ &  $79.59 \pm 2.41$ &  $81.52 \pm 2.36$ \\
            & distilroberta &  $89.14 \pm 1.19$ &  $83.93 \pm 2.76$ &  $79.47 \pm 1.98$ &  $81.35 \pm 1.93$ \\
\bottomrule
\end{tabular}

    \caption{Aggregated results of the tenfold cross-validation for all tested models. Best results for each metric are highlighted in bold.}
    \label{tab:model_selection}
\end{table*}

As shown in  ~\tabref{tab:model_selection}, the combination of CNN and RoBERTa not only reaches the highest average scores in accuracy and macro F1, but also the lowest standard deviation in the tenfold cross-validation. %
This leads us to believe that this model will not only generalize well but also will less likely deviate from the performance, which is why we choose this model as our classifier.
\begin{table}[h]
\small
    \centering
\begin{tabular}{llll}
\toprule
{} &                F1 &                 P &                 R \\
\midrule
0  &  $94.17 \pm 0.64$ &  $92.83 \pm 1.13$ &  $95.57 \pm 0.95$ \\
1  &  $59.62 \pm 7.28$ &  $64.68 \pm 9.55$ &  $56.15 \pm 8.38$ \\
2  &  $86.74 \pm 2.09$ &   $86.11 \pm 4.2$ &  $87.52 \pm 2.02$ \\
99 &  $88.72 \pm 3.09$ &  $95.69 \pm 4.19$ &  $82.94 \pm 5.04$ \\
\bottomrule
\end{tabular}

    \caption{Detailed per-class results of the cross-validation for the chosen model (CNN + RoBERTa).}
    \label{tab:detailed_results}
\end{table}
~\tabref{tab:detailed_results} shows detailed results for the final model per label. It shows that predicting the label 1 is most difficult. Label "99" can be predicted with an F1 value of around $83\%$. %
We follow the argumentation in Weikmans et al. (\citeyear{Weikmans2017AssessingTC}) and regard these predictions as Rio marker 0. %
The influence of these examples is negligible as, ultimately, 99 is predicted in only $<0.04\%$ of the CRS in the end. \\
\indent The Rio marker classification model is a proxy to identify overreported cases. %
We are interested in those cases where our classification algorithm differs from the reported Rio marker, specifically classifying lower than the reported value %
We define overreporting $o$ of activity $x$ as
\begin{equation}
\textbf{o(x)} =
\left\{
	\begin{array}{ll}
		1  & \mbox{if } \text{reported(x) }>\text{ classifier(x)} \\
	    0  & \mbox{otherwise} \\
	\end{array}
\right.  
\label{eq:overreported_definition}
\end{equation}
This leads to three cases of overreporting: The classifier predicts 0 and the Rio marker reports 1 or 2, and the much harder case where Rio marker reports 2 and the classifier predicts 1. %

\subsection{CARE Data}
 We apply our classifier to the CARE data set and compare the findings to the manual CARE annotations. %
 The set of examples in CARE data set are distinct from those in the WK data. %
We create an overreported flag for the CARE data by comparing the reported Rio marker to their re-evaluation markers using Equation~\eqref{eq:overreported_definition}. %
This marks $\caredataoverreported0$\% of the data as overreported. %
Our classifier predicts an overreporting rate of $\caredataoverreportedclassifier\%$ on the same data, indicating a significant difference in the annotation schemes. %
WK annotations appear stricter, leading to higher rates of overreporting than the CARE re-evaluation.

\subsection{Extrapolation of CARE Annotation using the Bayesian Formula }
    Using the Bayesian formula, we estimate the difference in annotation scheme and extrapolate it. %
    As discussed above, our training relies on the WK data set. %
    The approach of training and classifying new data ultimately transfers their annotation scheme to other data sets. Given the high number of hand-coded aid activities, the WK data set is well suited for training purposes. %
    However, comparing the resulting classification with CARE data, we find that this might overestimate overreporting. %
    We estimate the probability that if the classifier would mark any sample as overreported according to the WK data annotations, CARE annotations would agree, and vice versa. %
    Using the Bayesian formula, we update the estimation of our classification. %
In mathematical formulation we define two events: $W$ (classifier marks sample as overreported) and $C$ (CARE annotates sample as overreported). %
We further denote the data set from which we calculate the corresponding term as the parameter $D$.
The Bayesian formula is then:
\begin{equation}
    P(C; D\text{=CRS}) = \frac{P(C|W; D\text{=CARE})}{P(W|C; D\text{=CARE})}\cdot P(W; D\text{=CRS})
    \label{eq:bayes}
\end{equation}
We note that this Bayesian formulation makes implicit assumptions about the independence of annotation schemes and data samples. We argue that since the CRS data is the basis for all of these samples, that these simplifications are acceptable and lead to a simple model to show the potential and benefits of this approach. We plan to investigate and apply more complex models to these dependencies in future work.

We calculate $P(W|C)$, i.e. the probability that our classifier would agree with CARE, and $P(C|W)$, i.e. the probability that CARE would agree with our classifier, from the CARE data set. %
Since the calculation is based on a small sample, we consider the uncertainty of the estimate using the beta distribution to approximate the factors and simulate the propagation of these uncertainties:
\begin{equation}
    P(W|C) \propto \text{beta}(1+n, 1+m),
\end{equation}
where $n$ is the number of positive examples and $m$ the number of negative examples in the data. We then report the credible interval of $95\%$.
This leads to the correction factor
\begin{equation}
   \frac{P(C|W)}{P(W|C)} = 42.57\%\, ([26.47\%; 64.39\%])\quad.
\end{equation}
We denote the credible interval of $95\%$ in brackets behind the point estimate.
Using the same procedure, we propagate the correction factor to adjust the overall overreporting rate using Equation ~\eqref{eq:bayes}. %

\subsection{Extrapolation and Exploration}
We can now apply the classification algorithm to the CRS data. %
We restrict the CRS to projects that have a Rio marker higher than 0, otherwise, by definition, they cannot be overreported in Equation ~\eqref{eq:overreported_definition} and we consider only at the top five DAC donors: France, Germany, Japan, the United Kingdom and the United States. %
We use the fastText \cite{joulin-etal-2017-bag} language detection to classify the language of the descriptions. %
While Germany, Japan, the United Kingdom and the United States report almost all their projects in English, France tends to report in French. %
The classifier was also trained on French descriptions from the WK set, however, we predict Rio markers for these projects using both the original French descriptions and also automated translations into English using Google Translate (the influence of which is discussed below). %
After that, the data set contains 46,280 projects from 2010 to 2019 with short and long textual descriptions and a reported Rio marker of 1 or 2. %
This also complies with how data was sampled in the WK and the CARE data sets. \\
\indent Figure \ref{fig:plot_classification} shows the results of the extrapolation. %
\tabref{tab:care-extrapolation} in Appendix \ref{app:overreporting_per_year} shows the underlying values and the number of observations. %
Extrapolation is done, again, by concatenating the project title and long description into a text string and feeding it into the network. %
The network assigns a Rio marker prediction to every project. %
After that we use Equation ~\eqref{eq:overreported_definition} to mark all activities with a flag for overreporting. 
The classifier detects an overall overreporting rate of \textbf{\overreportingcrsdata \%} in the CRS in terms of WK annotations and an estimated $\mathbf{32.03\%\, ([19.81\%; 48.34\%])}$) in terms of CARE annotation.

\subsection{The Influence of Input Length}
\label{sec:influence_text_length}

Systematic variation of text length by donor or year might bias our results. %
Longer descriptions usually contain more information and thus improve the validity of the classification. %
Elaborating on the difficulty of classifying short texts is beyond the scope of this paper and is its own established field of research, e.g. \cite{Chen2019DeepST,Wang2017CombiningKW}. 
This should specifically pertain to cases where CCA is a `significant' but not a `principal' objective (i.e., $r=1$). Here, descriptions might not mention CCA because it is not the main motivation of the aid activity. 
Moreover, it seems likely that very short descriptions are mostly classified as $r = 0$ for lack of information.%

\begin{figure*}[ht]
    \centering
    {\includegraphics[width=1\textwidth, 
    trim={0cm 0cm 0cm 0cm} , clip]{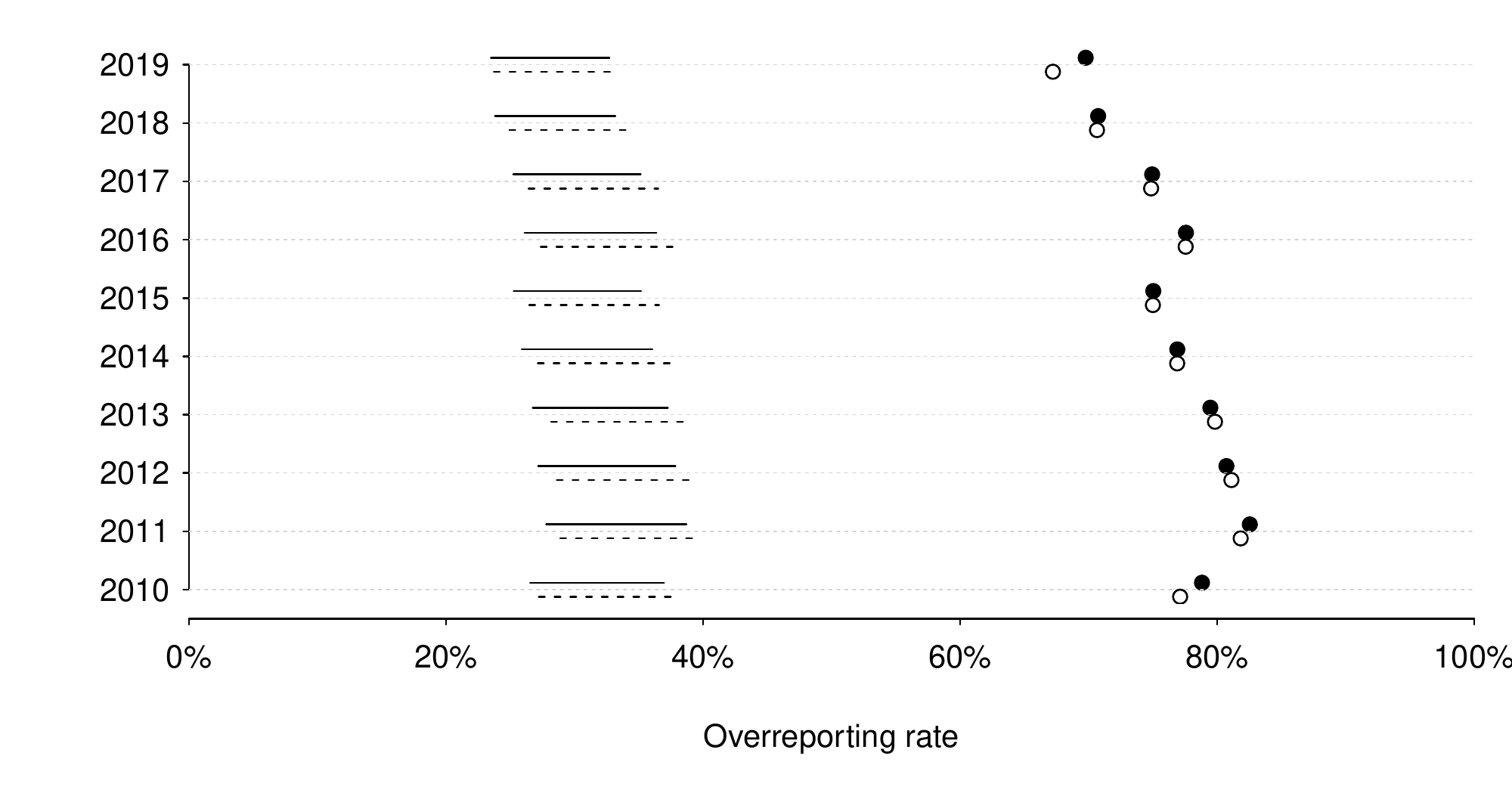}}
    \caption{Estimated overreporting rate extrapolation of WK data by year (dots) and estimated correction based on CARE data (lines). Black dots and solid lines show estimates based on the full sample. Hollow circles and dashed lines show estimates based on samples excluding very short descriptions.}
    \label{fig:plot_classification}
\end{figure*}

\begin{figure*}[ht]
    \centering
    {\includegraphics[width=1\textwidth, trim={0cm 0cm 0cm 0cm} , clip]{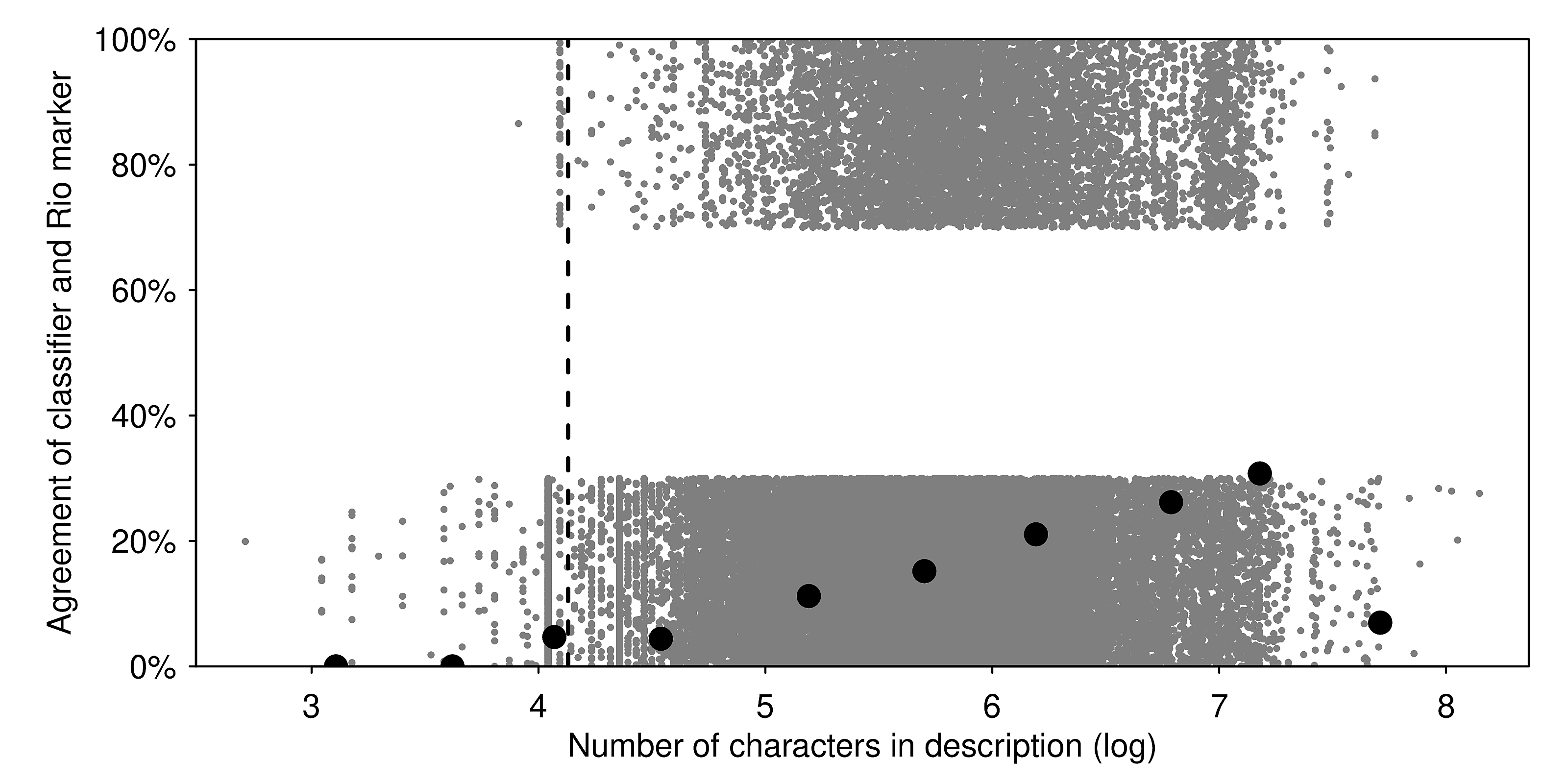}}
    \caption{Agreement of classifier with Rio marker by lengths of activity description. Random noise is added to the point locations in the y-direction for better visibility. Circles show averages of binned data. The vertical dotted line indicates the cut-off for outliers based on the IQR method for outliers.}
    \label{fig:plot_agree_lengths}
\end{figure*}

We find evidence that classification quality correlates with description length.
As we do not know the true Rio marker for aid activities, we use the rate of agreement between our classification and the assigned Rio marker as an indicator of classification quality.
We assume that very low rates of agreement indicate poor performance of classification. 
As Figure~\ref{fig:plot_agree_lengths} shows, the share of cases where classification results and Rio marker are identical increases with description length. \\
\indent Description lengths systematically differ by donor and year (for both: p < 0.00, Kruskal-Wallis rank sum test), although the distribution of description lengths across donors and years shows that, overall, absolute differences are not large. 
Regarding donors, Japan is an exception, with considerably shorter descriptions than the median of the other donors (78 vs. 315 characters, respectively).
The time series shows an increasing trend where the median of description length increased from 231 to 345 characters between 2010 and 2019 (see also Appendix B, Figure~\ref{fig:plot_lengths_by_donor_year}). 
If classification quality depends on description lengths, and description lengths vary by donor and year, this could introduce confounding bias distorting the comparison of overreporting rates across years and donors. 
More specifically, an increase in description lengths could be interpreted erroneously as a decrease in overreporting. 
As an example, Japan's high rates of overreporting (see Figure~ \ref{fig:plot_per_donor_per_year}) could be partly explained by the brevity of aid descriptions in the CRS. \\
\indent To account for possible distortion of our results, we rerun our analysis excluding short descriptions from our estimation of overreporting.
We used the interquartile range (IQR) method to identify outliers \citep[p. 12]{ilyas_data_2019}, i.e. we excluded descriptions with lengths below the threshold of $Q1 - 1.5 \times IQR$ (in our case, 62 characters; logarithmic: $4.1$). %
This seems appropriate as indicated by an increase in agreement between Rio marker and classifier (see Figure~\ref{fig:plot_agree_lengths}). \\
\indent We rerun the analysis that created ~\tabref{tab:care-extrapolation}, but excluded all projects with fewer than 62 characters. We also excluded from the CARE data set  data points shorter than 62 characters, when calculating the correction factor. A comparison of the results is presented in Figure ~\ref{fig:plot_classification}. %
The overall overreporting rate per year drops slightly, while the estimation based on CARE increases. %
This stems from the fact that the classifier agrees with the high-quality CARE re-evaluation more often for longer texts. The correction factor and uncertainty in this case slightly increases from 42.57\% ([26.47\%; 64.39\%]) to 44.32\% ([27.55\%; 66.62\%]).%
Overall this leads to slightly lower overreporting rate according to WK but a slightly higher estimate for CARE-corrected overreporting.

In summary, we do find evidence that systematic variation of text length by donor or year can bias results. %
However, robustness tests indicate that the bias does not change overall conclusions. 

\subsection{Extrapolation Per Donor Per Year}
\begin{figure*}[ht]
    \centering
    {\includegraphics[width=1\textwidth, trim={2.4cm 1.1cm 2.9cm  1.7cm},clip]{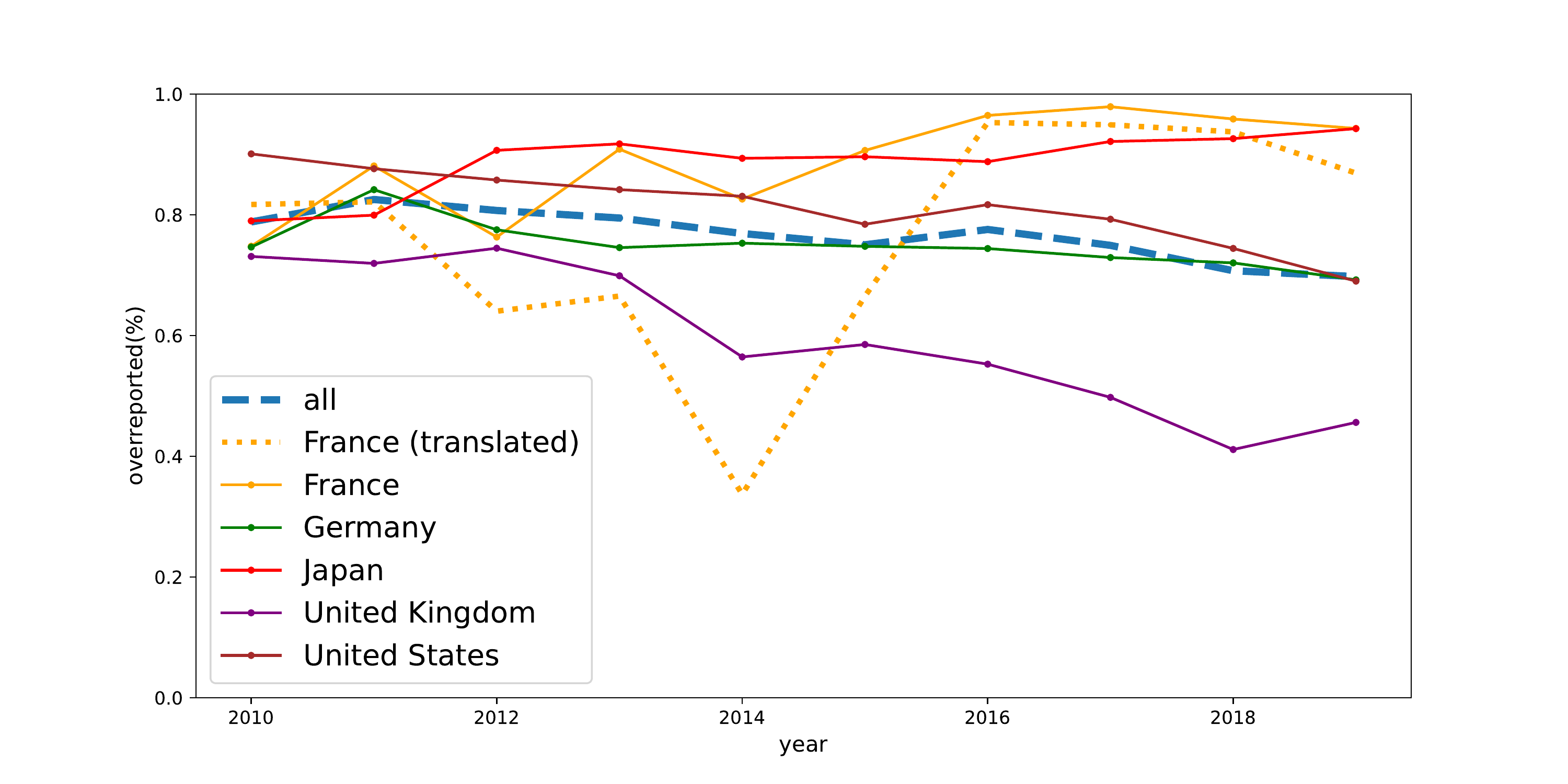}}
    \caption{Rates of overreporting  by donor country, 2010 to 2019}
    \label{fig:plot_per_donor_per_year}
\end{figure*}

\begin{table}[ht]
\centering
\small
\begin{tabular}{llllll}
\toprule
Year &                       France &                      Germany &                        Japan &               U.K. &                U.S. \\
\midrule
2010 &   \cellcolor{Red!50}$74.8$\% &  \cellcolor{Red!27}$74.66$\% &  \cellcolor{Red!24}$78.99$\% &  \cellcolor{Red!18}$73.11$\% &  \cellcolor{Red!15}$90.09$\% \\
2011 &   \cellcolor{Red!83}$88.1$\% &   \cellcolor{Red!0}$84.16$\% &  \cellcolor{Red!30}$79.94$\% &   \cellcolor{Red!1}$71.95$\% &   \cellcolor{Red!0}$87.63$\% \\
2012 &  \cellcolor{Red!77}$76.32$\% &   \cellcolor{Red!0}$77.52$\% &  \cellcolor{Red!12}$90.68$\% &   \cellcolor{Red!0}$74.47$\% &   \cellcolor{Red!0}$85.76$\% \\
2013 &  \cellcolor{Red!47}$90.87$\% &   \cellcolor{Red!0}$74.57$\% &   \cellcolor{Red!0}$91.75$\% &    \cellcolor{Red!0}$69.9$\% &   \cellcolor{Red!0}$84.17$\% \\
2014 &  \cellcolor{Red!63}$82.61$\% &    \cellcolor{Red!0}$75.3$\% &   \cellcolor{Red!9}$89.36$\% &   \cellcolor{Red!8}$56.46$\% &   \cellcolor{Red!0}$83.07$\% \\
2015 &  \cellcolor{Red!48}$90.66$\% &   \cellcolor{Red!0}$74.78$\% &  \cellcolor{Red!15}$89.62$\% &   \cellcolor{Red!0}$58.52$\% &   \cellcolor{Red!0}$78.43$\% \\
2016 &   \cellcolor{Red!0}$96.46$\% &   \cellcolor{Red!0}$74.41$\% &   \cellcolor{Red!25}$88.8$\% &   \cellcolor{Red!0}$55.26$\% &   \cellcolor{Red!0}$81.68$\% \\
2017 &   \cellcolor{Red!0}$97.91$\% &   \cellcolor{Red!0}$72.92$\% &   \cellcolor{Red!0}$92.15$\% &   \cellcolor{Red!0}$49.76$\% &   \cellcolor{Red!0}$79.27$\% \\
2018 &   \cellcolor{Red!0}$95.87$\% &   \cellcolor{Red!0}$72.04$\% &   \cellcolor{Red!0}$92.62$\% &   \cellcolor{Red!0}$41.12$\% &   \cellcolor{Red!0}$74.44$\% \\
2019 &   \cellcolor{Red!0}$94.31$\% &   \cellcolor{Red!0}$69.23$\% &   \cellcolor{Red!0}$94.28$\% &   \cellcolor{Red!0}$45.61$\% &   \cellcolor{Red!0}$69.01$\% \\
\bottomrule
\end{tabular}

\caption{Overreporting rates - extrapolation of WK data split by donor country. Note that colored cells have fewer than 500 data points. Darker color corresponds to fewer data points.}
\label{tab:table_donor_per_year}
\end{table}%
The values in ~\tabref{tab:table_donor_per_year} show overreporting as identified by the classifier based on the WK data set from 2010 to 2019. %
France had the lowest number of projects in CRS reported with at least Rio marker 1 in 2011 and 2012 with 84 and 114 projects respectively. %
The rates in ~\tabref{tab:table_donor_per_year} for 2012 coincide rather well with the findings in \cite{Weikmans2017AssessingTC} for Germany ($81\%$), Japan ($92\%$) and the United Kingdom ($82\%$). %
The United States was not part of the sample in \cite{Weikmans2017AssessingTC}.\\
\indent The classifier detects significantly lower overreporting rates for France ($76.32\%$) than in the original paper ($92\%$) in 2012. %
We found that, of the 114 projects reported in 2012 by France, only 58 have unique descriptions, which produces very high uncertainty. %
Manual evaluation shows that there are cases of the same description occurring up to 11 times and is marked as not overreported. %
This alone can account for around a $10\%$ difference in overreporting measure when classified differently. %
We therefore marked the fields where there are a fewer data points than 500 in orange in ~\tabref{tab:table_donor_per_year} to show where these effects could have larger impact. \\
\indent Figure \ref{fig:plot_per_donor_per_year} also shows the classification results when translating the French descriptions to English (dotted line). %
There are significant differences in overreporting rates in the period 2013-2015, while the two lines are reasonably close between 2016 and 2019. 
The year 2014 in particular shows a difference of around $50\%$, from $33.7\%$ in the translated case to $82.6\%$ when using the original French descriptions. %
We argue that this happens for two reasons: %
First, the years with the biggest differences when translating descriptions coincide with the years when very few projects reported, thus the influence of a single misclassification is higher. Second, there is considerable noise in the data, which further increases variance in prediction when switching language. 

\begin{table}[ht]
\centering
\small

\begin{tabular}{p{1cm}p{1cm}p{1.5cm}rrp{1.2cm}}
\toprule
{} & Over-reported & Over-rerported (translated) & Count & Unique & Unique (relative) \\
year &               &                             &       &        &                   \\
\midrule
2010 &        74.80\% &                     81.71\% &   246 &    133 &           54.07\% \\
2011 &        88.10\% &                     82.14\% &    84 &     80 &           95.24\% \\
2012 &       76.32\% &                     64.04\% &   114 &     57 &            50.00\% \\
2013 &       90.87\% &                     66.54\% &   263 &    186 &           70.72\% \\
2014 &       82.61\% &                      33.70\% &   184 &     53 &            28.80\% \\
2015 &       90.66\% &                     66.54\% &   257 &     89 &           34.63\% \\
2016 &       96.46\% &                     95.29\% &   594 &    171 &           28.79\% \\
2017 &       97.91\% &                     94.91\% &  1100 &    477 &           43.36\% \\
2018 &       95.87\% &                     93.74\% &   847 &    372 &           43.92\% \\
2019 &       94.31\% &                     86.91\% &  1054 &    669 &           63.47\% \\
\bottomrule
\end{tabular}

\caption{Overreporting for France following the WK data set by years. The second and third columns denote the estimated overreporting rate for French descriptions and English translations, respectively. The following columns show: number of projects, number of unique descriptions, and the ratio of these two.}
\label{tab:french_language}
\end{table}%

In 2014, France reported 184 projects, but there are only 53 unique descriptions. %
The most frequent description was used 36 times, each of the occurrences having a unique ID in the CRS. %
Half of these 36 projects are reported with a Rio marker 1 and half with Rio marker 2. %
Every classifier predicting a Rio marker on the basis of these descriptions will therefore differ from the reported value at least half the time. %
This alone accounts for a difference in overreporting of $10\%$ in that year. %
This argument also holds for the second and third most frequent project descriptions, which were used 19 and 18 times respectively. %
When considering only the unique descriptions, the predictions' detection of overreporting in English and French agree in $77\%$ of cases, while only in $51\%$ of cases overall. %
In general, the larger the number of projects, the smaller the influence of this phenomenon. %
However, of the projects that France reported, only around $40\%$ of the descriptions are unique (see ~\tabref{tab:french_language} for more details).

\section{Discussion}
Our re-evaluation of aid activities reported as contributing to CCA indicates a lack of quality in the self-reporting of donors. %
A substantial share of reported adaptation aid activities does not explicitly mention CCA in project descriptions. %
This is problematic because valid indicators are required to assess whether the international community is meeting its climate policy obligations as described in, for example the Paris Accord. %
\indent Our finding indicates an overestimation of adaptation aid. %
Even after downward adjustment of our estimates to account for insufficient information from short project descriptions, our best estimate suggests that about every third activity categorized by donors as adaptation aid is not adaptation related. %
However, we cannot say whether this is due to a lack of clear reporting standards \cite{weikmans_what_2020}, a lack of compliance with reporting standards, or even incentives to report more than is actually delivered \cite{MICHAELOWA20112010}. %
Moreover, our estimates are subject to considerable uncertainty because an unambiguous classification of aid activities based on the Rio Marker methodology requires extensive knowledge of individual aid activities. \\
\indent Although the estimates are somewhat uncertain, our results confirm earlier findings of a substantial discrepancy between the figures reported by donors and re-evaluations by independent researchers \cite{MICHAELOWA20112010, Weikmans2017AssessingTC,junghans_different_nodate,Schramek2021ADAPTATIONF}. %
This can be partly explained by the fact that our approach draws on earlier classifications as training data and is thus not completely independent. %
However, our study also goes well beyond existing research in temporal and geographical scope: %
We assess every adaptation aid activity reported by OECD DAC donors since the Rio marker on adaptation aid was introduced in 2010. %
If donors had changed their reporting practice, we would likely see this in our data, yet, overall, we find no indication that reporting practices have changed significantly since 2010. %
The share of overreporting remains at a high level between 2010 and 2019. Although the classifier indicates a slight decrease of overreporting, the fluctuations are within the range of uncertainty of our adjusted estimates. %
Nevertheless, the results are mainly driven by a significant decrease in overreporting by the United Kingdom. %
However, we are careful to infer from our data substantial differences in reporting standards between countries; %
Country-specific results should be examined more closely in future research. %
We have no reason to suspect that the classifier has particular problems with data reported by the UK based on the results of the cross-validation and given the fact that most project descriptions are in English.
\section{Conclusion}
In this paper we propose an automated way of detecting overreporting of climate adaptation finance based on CRS project descriptions. %
Our approach is based on state-of-the-art text classification using finetuning neural language models. %
 We consider the quality of the annotations of our training data when estimating overall overreporting rates, and propose a Bayesian approach to estimate an extrapolation of high-quality annotations. %
Our approach indicates significant overreporting throughout the study period. 

 There are two key challenges with this approach: %
 The quality and quantity of annotation scheme data, and the quality of the textual input. %
 While, ultimately, the first challenge can be overcome by extending data with higher-quality annotations, the second proves trickier. %
 Unfortunately, the CRS data does not have uniform quality built into its textual descriptions. %
 We have discussed the influence of description length and language on the quality of our classifier, but while future work may incorporate techniques from short-text classification research, in many cases the information the CRS contains will likely not suffice to use techniques like augmentation or conceptualization. %
 To improve on this, additional external data would be necessary, to which, at this point, we had no access. %
 Especially deciding if the CCA aspects of a project comprises a "significant" or "principal" object should benefit from this. %
 Another way to improve would be to pay more attention to multi lingual classification research and either incorporate techniques for multi lingual text classification or utilize a high-quality pipeline for translation.

\newpage

\bibliographystyle{plainnat}

\appendix
\newpage
\newpage

\section{Overreporting per Year}
\label{app:overreporting_per_year}
Tables \ref{tab:care-extrapolation} and \ref{tab:care-extrapolation-n62} show detailed numerical results on estimating overreporting split by year. These two tables were summarized in Figure \ref{fig:plot_per_donor_per_year}, comparing the resulting rates.

\begin{table}[ht]
\small
\centering
\begin{tabular}{lllr}
\toprule
{} & Classifier &   CARE Estimated &    count \\
year &            &                  &          \\
\midrule
2010 &  $77.06$\% &  $20.31-49.46$\% & $1683$\% \\
2011 &  $81.76$\% &  $21.55-52.48$\% & $3076$\% \\
2012 &   $81.0$\% &  $21.35-51.99$\% & $2858$\% \\
2013 &  $79.75$\% &  $21.02-51.18$\% & $3698$\% \\
2014 &  $76.79$\% &  $20.24-49.28$\% & $3739$\% \\
2015 &  $74.91$\% &  $19.75-48.08$\% & $4308$\% \\
2016 &  $77.49$\% &  $20.43-49.73$\% & $5300$\% \\
2017 &  $74.78$\% &   $19.71-48.0$\% & $7284$\% \\
2018 &   $70.5$\% &  $18.58-45.25$\% & $6631$\% \\
2019 &  $66.99$\% &   $17.66-43.0$\% & $7111$\% \\
\bottomrule
\end{tabular}

\caption{Overreporting rates - extrapolation of WK data and the estimated correction of CARE data.}
\label{tab:care-extrapolation}
\end{table}%
\begin{table}[ht]
\small
\centering
\begin{tabular}{lllr}
\toprule
{} & Classifier &   CARE Estimated &    count \\
year &            &                  &          \\
\midrule
2010 &  $78.77$\% &   $21.7-52.48$\% & $1818$\% \\
2011 &  $82.48$\% &  $22.72-54.95$\% & $3207$\% \\
2012 &  $80.62$\% &  $22.21-53.71$\% & $2894$\% \\
2013 &  $79.39$\% &  $21.87-52.89$\% & $3731$\% \\
2014 &   $76.8$\% &  $21.16-51.16$\% & $3745$\% \\
2015 &  $74.94$\% &  $20.65-49.93$\% & $4314$\% \\
2016 &  $77.52$\% &  $21.36-51.64$\% & $5307$\% \\
2017 &  $74.86$\% &  $20.62-49.87$\% & $7306$\% \\
2018 &  $70.59$\% &  $19.45-47.03$\% & $6654$\% \\
2019 &  $69.56$\% &  $19.16-46.34$\% & $7710$\% \\
\bottomrule
\end{tabular}

\caption{Overreporting rates - extrapolation of WK data and the estimated correction of CARE data after eliminating all inputs with fewer than 62 characters.}
\label{tab:care-extrapolation-n62}
\end{table}%

\section{Analysis of Text Length}
\label{app:analaysis_of_text_length}
\begin{figure*}[ht]
    \centering
    {\includegraphics[width=0.75\textwidth, 
    trim={0cm 0cm 0cm 0cm} , clip]{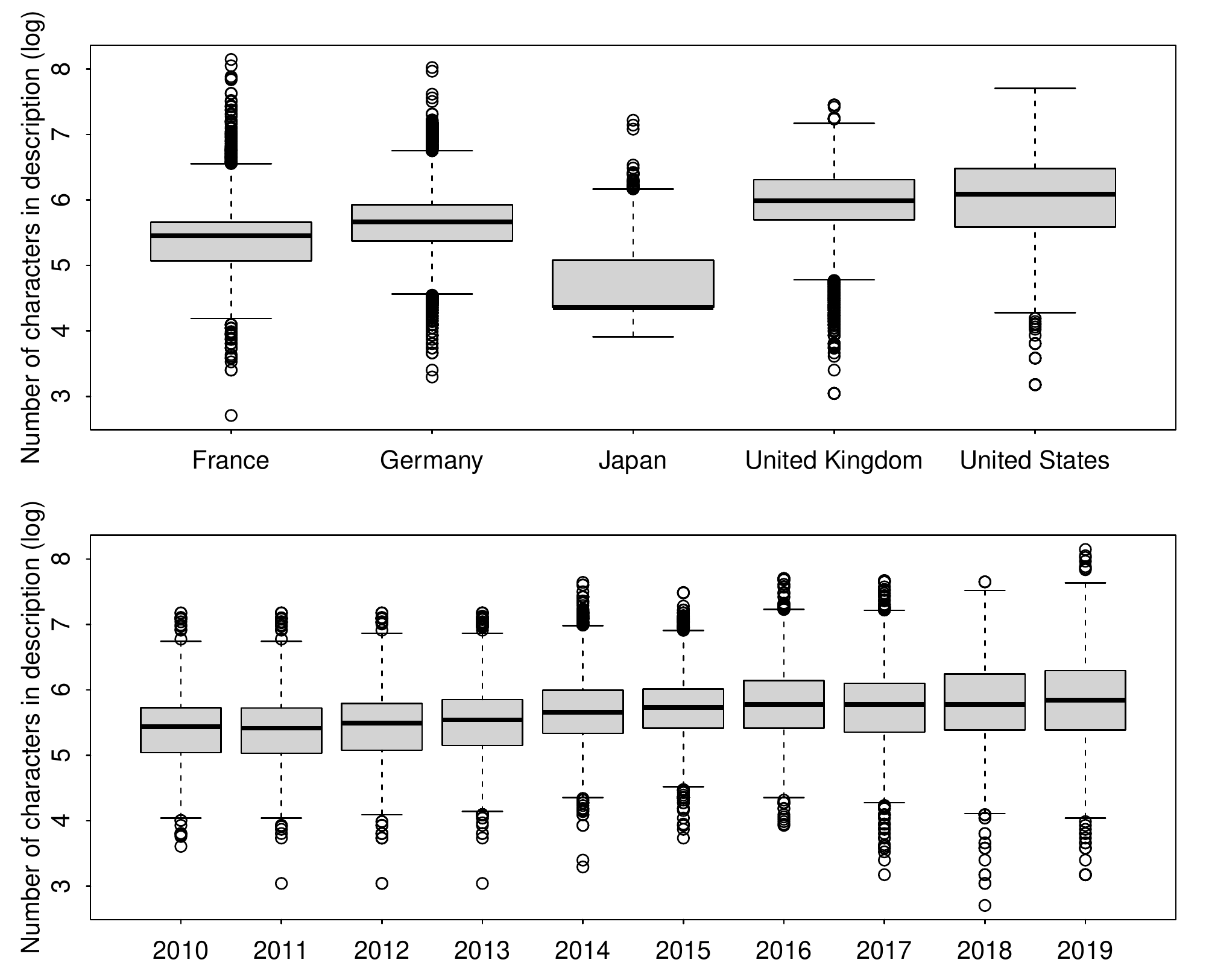}}
    \caption{Activity description lengths (logarithmic) by donor (top figure) and by years (bottom figure).}
    \label{fig:plot_lengths_by_donor_year}
\end{figure*}

Figure \ref{fig:plot_lengths_by_donor_year} illustrates increasing average descriptions lengths over the study period in a standard box plot, for the argument in section \ref{sec:influence_text_length}. %
Also, Japan tends to use fewer characters in descriptions compared to the other donors. %
This supplements Figure \ref{fig:plot_agree_lengths} and informs the decision to use the \textit{IQR} range to quantify the lower limit for the number of characters. %
This also checks that cutting off at 62 characters (logarithmic: 4.1) does not introduce a bias for a particular year or donor.

\end{document}